\documentstyle[12pt,epsfig]{article}
\begin{document}

\centerline{ Woven Geometries: Black Holes }
\vspace{0.3cm}
\centerline{\rm Junichi Iwasaki}
\centerline{ Physics Department, University of Pittsburgh,
                  Pittsburgh, PA 15260, USA}
\centerline{\rm E-mail:\ iwasaki@phyast.pitt.edu}
\vspace{0.2cm}
\centerline{December 12, 1994}
\begin{abstract}
Tangles of loops which approximate an aspect of the Kerr-Newman
black hole metrics at large scales compared to the Planck length
are constructed.
The physical aspect the tangles approximate is discussed.
This construction may be useful in the loop representation of
canonical quantum gravity.
Implications and applications of the tangles are remarked.
\end{abstract}

\section{Introduction}\label{sec:intro}

It is known in the canonical formalism that the gauge field
admits loop-wise excitations,
if no external source is present,
due to the gauge invariance requirement
in contrast to  non-gauge fields
(i.e. scalar and Dirac fields), in which
point-wise excitations are allowed.
This means that a solution of the classical equation
of the gauge field
could be thought as a sum of loops along which
the gauge potentials are ``concentrated".
For the Maxwell theory in a flat space-time, for instance,
the loop-wise excitations could be interpreted
as lines of electric force in the absence of charged particles.
A quantum description of the theory in terms of lines of force
instead of the traditional description in terms of particles
(photons) is possible \cite{Maxwell,FKNR}.

General Relativity (GR) has been reformulated
in the canonical formalism as a gauge field theory
in terms of connection variables \cite{Ashtekar}
and hence suggesting the existence of a description
in terms of loop-wise excitations.
Then the loop representation has been applied to
quantizing the theory \cite{loop}.
In this representation,
the geometrical information is coded in loops;
a quantum state is a functional of
loops $\gamma:S^1\to\Sigma$, where $\Sigma$ is a 3-space.
An advantage of the use of loops in GR is that the diffeomorphism
invariance, which is a key notion in GR, seems more tractable
than with the use of metrics.
In fact, diffeomorphism invariant quantum states have been realized
as functions of knot and link classes of loops in $\Sigma$,
namely ``knot states." \cite{loop}
In this respect relations between Quantum Gravity in physics
and Knot theory in mathematics have been revealed.
(For recent account, see \cite{baez}.)
A problem is the interpretation of such loops.
If one interprets them as lines of gravitational force, then
it is unclear how they describe classical space-time geometries,
which are responsible to the gravitational force in GR.

Some quantum operators (i.e. the area operator),
which have 3-metric information
classically, have been constructed and shown to be diagonal
in the loop representation \cite{weave}.
This fact could suggest
that the loops more or less represent 3-metric geometries
and some sets of loops representing a certain aspect of 3-metric
geometry may exist.
Tangles of loops, called ``weaves,"
which approximate known 3-metrics in the following sense
have been constructed;
the areas of a given surface
measured on a weave state and its corresponding metric state
coincide up to an error of order of $l_p/L$,
where $l_p$ is the Planck length and $L$ is the scale of
the measurements, provided that the surface is slowly varying
over $L$ with respect to the metric corresponding to the state.
A hope could be that each of an infinite number of different weaves,
a sector of the entire loop basis, might approximates one of an
infinite number of different metrics and hence the loop basis
might ``generalize" the notion of metric geometry.
Moreover, the loops might provide an aspect of ``precise"
picture of the Planck scale geometry, to which classical
picture of geometry is insensitive, and
induce 3-metrics at large scales compared to the Planck length.

However, in the constructions,
the weaves were determined from classical metrics,
hence they could not contain more information than
the corresponding metrics.
If the loop picture is correct, then ultimately any metric should
be determined from a weave and hence the weaves should be more
``precise" than the corresponding metrics.
At present we know only small number of weaves determined
from classical metrics:
the weaves representing flat metrics\cite{weave},
Schwarzschild metrics (outside the horizon)\cite{joost}
and gravitational plane wave metrics \cite{Borissov}.

Before exploring the ``precise" structure of space-time
in terms of loops, if it exists,
one may need to understand at least a mechanism of determining
a weave from a given metric,
since one is familiar with only the notion of metric in classical GR.
In order to understand the mechanism and
how the loops represent metrics in a physically viable way,
it could be useful to consider metrics with more physically meaningful
parameters and metrics to which quantum effects are supposed to
be important.
In this paper, we construct weaves representing the Kerr-Newman
black hole metrics,
a three parameter family of solutions of the Einstein equation
and known to be the only stationary black holes.
Of course, our construction here by no mean exhausts
the understanding of the mechanism.
Rather, it should be understood as a step toward an unseen goal.
In Sec.~\ref{sec:weave} we review the reference \cite{weave}
about the idea of the weaves and a way of constructing weaves
representing curved metrics from a flat weave.
In Sec.~\ref{sec:kerr-new} we construct weaves representing
the Kerr-Newman black hole metrics and discuss physical information
containd in the weaves.
In Sec.~\ref{sec:remarks} we give some remarks
on implications and applications of the weaves constructed
in this paper.

\section{Flat and curved weaves}\label{sec:weave}

The bra states in the loop representation of quantum gravity
labelled by countable loops are analogous to the countable
energy-angular-momentum states of the hydrogen atom
labelled by a set of integers $(n,l,m)$.
However, our understandings of the two cases are different.

In the latter case, our understanding of the states is complete
in the sense that we know the correct inner product and hence
we have the corresponding ket states
and these bra and ket states determine
the quantum dynamics of the hydrogen atom.
Rather than specifying the position of the electron
as a function of time, a set of three integers $(n,l,m)$
is essential to specify the entire physical information of
the hydrogen atom.

In the former case, our understanding of the states is incomplete
in the sense that we do not know a correct inner product
and hence we do not have ket states.
We hope that rather than specifying the 3-metric $q_{ab}$
of space as a function of (coordinate) time,
a countable number of loops is essential to specify
at least an aspect of physical information of
space geometry.
We do not expect that the loop basis determines the
entire physical information of the dynamics of gravity.
Rather we would like to use them as an ansatz
to investigate the quantum dynamics of gravity
in contrast to the case of the hydrogen atom,
in which the energy-angular-momentum
basis is a final result of quantum mechanics.
Note that this ansatz is not an assumption but a result \cite{weave}
in the kinematical level, namely before imposing
the constraint conditions,
by which the dynamics of gravity is induced.

For this strategy to be physically viable,
it is desired to understand the physical
meaning of the loops and their relation to the traditional
description of space-time in terms of metrics.
Since some quantum operators which have 3-metric information
classically
have been shown to be diagonal in the loop basis \cite{weave,discrete},
it is natural to expect that the loop states more or less
represent 3-metric geometries.

An idea which relates loops to metrics is the construction of
the weaves.
A flat weave is a tangle of loops which approximates an aspect
of a flat metric at large scales relevant in classical GR.
The aspect is determined by a set of operators one is interested in.
So far the area operator has been used and a resulting flat weave
approximates the flat metric in the sense that the areas of
a given surface measured on the weave state and its corresponding
metric state coincide up to a error of order of $l_p/L$,
where $L$ is the scale of the measurements.
There are still an infinite number of
weaves sharing this aspect of the metric.
If one imposes more conditions by means of more operators,
then one might determine a more restricted set of weaves
reflecting a wider range of aspects of the metric.
Within the restricted set of weaves,
whether one particular weave or a linear combination of the weaves
represents the classical metric has not been understood.
The flat weave we make use of is a weave determined
by the area operator only.
Furthermore, it is a particular choice among an infinite number
of the weaves sharing the same aspect determined by the area operator.
The curved weaves constructed by deforming the flat weave
also reflect the same aspect.

A way of constructing a flat weave is the following.
Given a flat 3-metric $h_{ab}$,
the flat weave $\Delta$ is a set of
randomly oriented circles of radius
$a$ whose
centers are randomly distributed in the 3-space with
an average number density $n=1/a^3$;
$a$ is of order of the Planck length $l_p$.
Some of these circles may be linked to or
may intersect with one another.
We call each circle (or its deformation) a loop component.
There is an infinite number
of different flat weaves $\Delta$,
and each one corresponds to a different
loop state
 $\langle\Delta|$, namely , a different point in the state space of
loops; these states are denoted as flat weave states.
If $S$ is a macroscopic surface whose extrinsic curvature varies
slowly at the scale $L$,
then the eigenvalue of the area operator acting on any of
the flat weave states agrees with the area of $S$
with respect to the flat metric
up to an error of order $l_p/L$, that is,
\begin{equation}
l_p^2\int_S\left\vert\oint ds\dot\Delta^a(s)\delta^3(x,\Delta(s))
dS_a(x)\right\vert
=\int_S \left(\tilde{\tilde h}^{ab}dS_a(x)dS_b(x)\right)^{1/2}
+{\cal O}(l_p/L),
\end{equation}
up to a multiplicative constant in the left hand side,
where $\tilde{\tilde h}^{ab}$ is the double densitized
inverse of $h_{ab}$ and $dS_a(x)$ is
an infinitesimal surface element at $x$.
It has been shown by rigorously regularized calculations
\cite{weave,discrete}
that the value is proportional to the number of times
the flat weave $\Delta$ intersects the surface $S$.
In this respect all the flat weave states approximate equally well the same
flat metric for large enough $L$.

In addition, another significant result from the consideration
of the weave state is that the geometrical structure at the Planck scale
is discrete (consisting of countable loops)
due to the requirement that $a$ is a finite number of order $l_p$,
because of which the weave
approximates a classical continuum geometry
at large scales \cite{weave}.
The weave could be seen as a ``classical" limit
of quantum geometry.

One can construct a curved weave by deforming a flat weave.
If one define a tensor $t^a_{\ b}$ such that a given curved metric
$q_{ab}$ is expressed by $q_{ab}=h_{cd}t^c_{\ a}t^d_{\ b}$ and
define a curved weave $\Delta_t$ such that
the coordinates $\Delta_t^a$ of each point on the weave is
$\dot\Delta^a_t=t(t^{-1})^a_{\ b}\dot\Delta^b$,
where $\Delta$ is a flat weave and $t$ and $(t^{-1})^a_{\ b}$
are the determinant and the inverse of $t^a_{\ b}$
respectively, then
\begin{eqnarray}
&&l_p^2\int_S\left\vert\oint ds\dot\Delta^a_t(s)
\delta^3(x,\Delta_t(s))dS_a(x)\right\vert=
l_p^2\int_S\left\vert\oint ds\dot\Delta^b(s)
\delta^3(x',\Delta(s))t(t^{-1})^a_{\ b}dS_a(x)\right\vert
\nonumber\\
&&=\int_S\left(\tilde{\tilde h}^{cd}t(t^{-1})^a_{\ c}
t(t^{-1})^b_{\ d}dS_a(x)dS_b(x)\right)^{1/2}+{\cal O}(l_p/L)
\nonumber\\
&&=\int_S\left(\tilde{\tilde q}^{ab}
dS_a(x)dS_b(x)\right)^{1/2}+{\cal O}(l_p/L),
\end{eqnarray}
where $x'$ is the position transformed from $x$ by $t(t^{-1})^a_{\ b}$.
This means that the eigenvalue of the area operator acting
on the weave state $\langle\Delta_t|$ agrees with the area of $S$ with
respect to the metric $q_{ab}$
($\tilde{\tilde q}^{ab}$ is its double densitized inverse).
In other words, the area of $S$ is the number of times the weave
$\Delta_t$ intersects the surface and its value agrees with
the value of the area of $S$ measured with respect to the metric
$q_{ab}$.
In this sense the weave $\Delta_t$ represents the curved metric.
This construction of a curved weave works where $q_{ab}$
is slowly varying with respect to $h_{ab}$.
We use this technology to construct weaves representing
the Kerr-Newman black hole metrics in the next section.

\section{Weaving black holes}\label{sec:kerr-new}

A Kerr-Newman metric with total mass $M$,
electric charge $e$
and angular momentum per unit mass $b=|J|/M$
in geometrized units is \cite{wald}
\begin{equation}
ds^2=-{\Lambda\over\Sigma}\left(dt-b\sin^2\theta d\phi\right)^2
+{r^4\sin^2\theta\over\Sigma}
\left({r^2+b^2\over r^2}d\phi-{b\over r^2}dt\right)^2
+{\Sigma\over\Lambda}dr^2+\Sigma d\theta^2,
\label{eq:kerr}
\end{equation}
where
$\Sigma=r^2+b^2\cos^2\theta$ and
$\Lambda=r^2+b^2+e^2-2Mr$.
It is known that, if $b^2+e^2\le M^2$,
this metric has an event horizon at
$r_+=M+\sqrt{M^2-b^2-e^2}$
and a complicated global structure
with an infinite number of black and white holes
and asymptotically flat regions,
which can be explored by extending
space-time through the coordinate singularities,
where $\Lambda=0$.
However, here we are not interested in the entire structure
of space-time but interested in an asymptotically flat region,
where we are supposed to live, and a black (or white) hole
region, which are supposed to be being discovered by
our astronomers, if it exists in the universe.
In order to extend space-time into a black (or white)
hole region,
we need an appropriate coordinate system.
We choose namely ``ingoing" (or alternatively ``outgoing")
geodesic coordinates.
The change of coordinates such that
$dT=dt\mp\left({r^2+b^2\over\Lambda}-1\right)dr$
and
$d\Phi=d\phi\mp(b/\Lambda)dr$ \cite{chandra}
leads to
\begin{eqnarray}
&ds^2=&-dT^2+dr^2+\Sigma d\theta^2
+\left(r^2+b^2\right)\sin^2\theta d\Phi^2
\pm 2b\sin^2\theta dr d\Phi
\nonumber\\
&&+{2Mr-e^2\over\Sigma}
\left(\mp dT+dr\pm b\sin^2\theta d\Phi\right)^2.
\label{eq:metric}
\end{eqnarray}
Furthermore, for the convenience of our construction
of a weave, we adopt a ``cartesian" coordinate system \cite{chandra}
defined by
\begin{equation}
x=(r\cos\Phi\pm b\sin\Phi)\sin\theta,\ \ \
y=(r\sin\Phi\mp b\cos\Phi)\sin\theta,\ \ \
z=r\cos\theta.
\label{eq:coord}
\end{equation}
Then, our final form of the Kerr-Newman metric,
which has a Kerr-Shild form \cite{chandra}, is
\begin{eqnarray}
&ds^2=&-dT^2+dx^2+dy^2+dz^2
\nonumber\\
&&+{2Mr-e^2\over r^2+{b^2\over r^2}z^2}
\left\{\mp dT+{1\over r^2+b^2}
\left[r(xdx+ydy)\pm b(xdy-ydx)\right]
+{1\over r}zdz\right\}^2,
\label{eq:carte}
\end{eqnarray}
where $r$ is not the radial coordinate corresponding to
$\sqrt{x^2+y^2+z^2}$ but a function of
$x$, $y$ and $z$ satisfying
\begin{equation}
{x^2+y^2\over r^2+b^2}+{z^2\over r^2}=1.
\label{eq:r2}
\end{equation}
The metric is asymptotically flat
and its induced 3-metric $q_{ab}$ at a constant $T$
is positive definite except
at the true singularity where $\Sigma=0$
and a small region where
$r\le -M+\sqrt{M^2+e^2-b^2\cos^2\theta}$.
In this paper, we do not consider these places.
The lower (upper) signs
in the right hand side of
Eq~(\ref{eq:metric}) or (\ref{eq:carte})
lead to the black (white) hole
interpretation.
Note that for a fixed $r$ Eq~(\ref{eq:r2}) specifies an elliptical
surface and the surface with $r=r_+$ corresponds to the event
horizon.

Let us now focus on the 3-metric $q_{ab}$.
Define a tensor $t^a_{\ b}$ such that
$q_{ab}=h_{cd}t^c_{\ a}t^d_{\ b}$,
where $h_{ab}$ is a flat 3-metric and
the indices  take the value $x$, $y$ or $z$
($h_{ab}\equiv \delta_{ab}$ in our coordinate system).
The components of $t^a_{\ b}$ are
\begin{equation}
\left[\matrix{t^x_{\ x}&t^x_{\ y}&t^x_{\ z}\cr
              t^y_{\ x}&t^y_{\ y}&t^y_{\ z}\cr
              t^z_{\ x}&t^z_{\ y}&t^z_{\ z}\cr}\right]
=\left[\matrix{1&0&0\cr 0&1&0\cr 0&0&\alpha}\right]
\left[\matrix{-{XZ/\rho}&-{YZ/\rho}&\rho\cr
               {Y/\rho}&-{X/\rho}&0\cr
                X&Y&Z\cr}\right],
\label{eq:tensor}
\end{equation}
with
\begin{eqnarray}
&&X={rx\mp by\over r^2+b^2},{\ \ \ \ \ }
  Y={ry\pm bx\over r^2+b^2},{\ \ \ \ \ }
  Z={z\over r},
\nonumber\\
&&\rho=\left(X^2+Y^2\right)^{1/2}=
\left({x^2+y^2\over r^2+b^2}\right)^{1/2},{\ \ \ \ \ }
  \alpha=\left(1+
{2Mr-e^2\over r^2+{b^2\over r^2}z^2}\right)^{1/2}.
\end{eqnarray}
Note that the condition (\ref{eq:r2}) implies $X^2+Y^2+Z^2=1$.

\begin{figure}
\par
\centerline{\epsfig{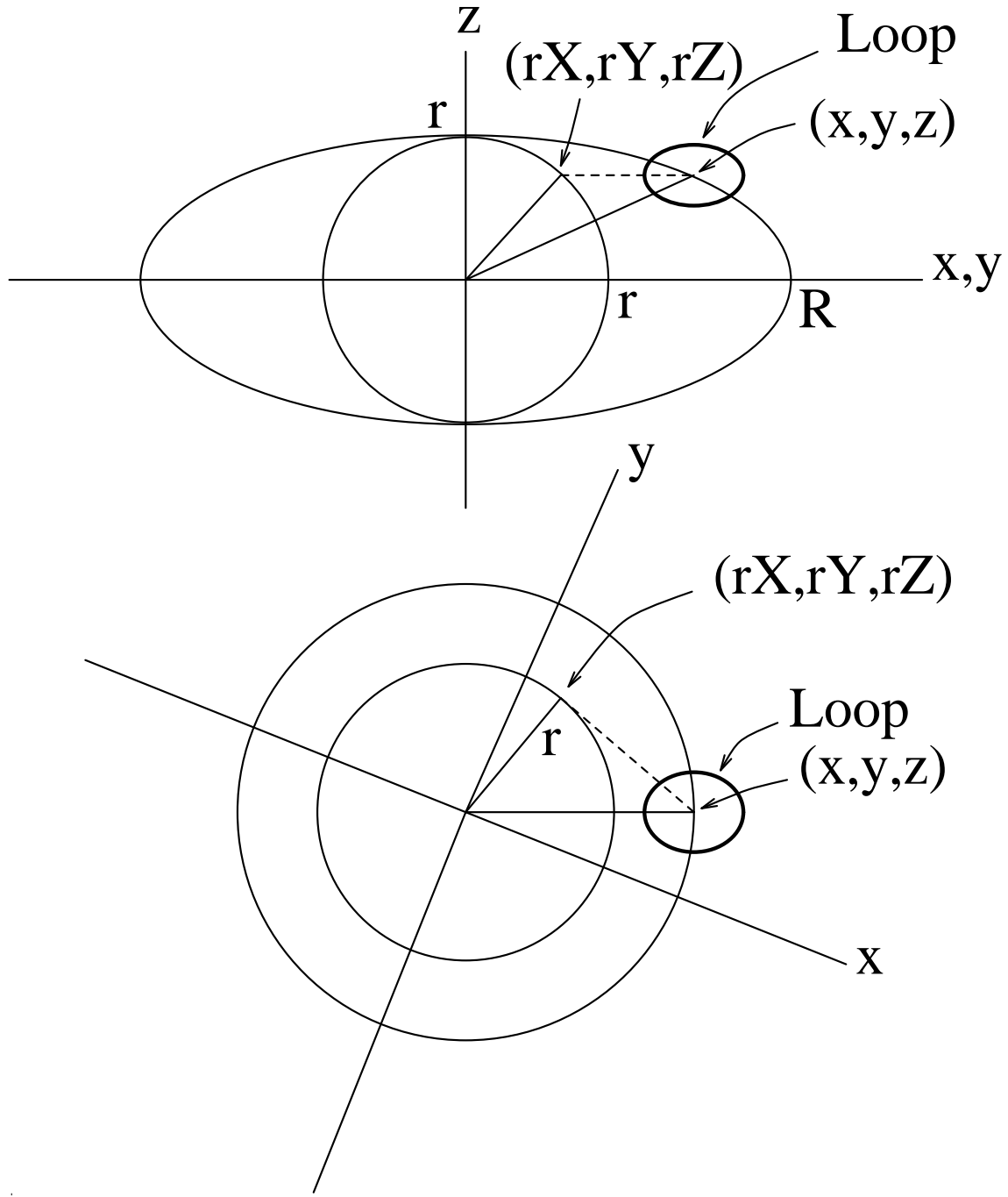}}
\par
\caption{A loop component at $(x,y,z)$.
For each $(x,y,z)$, there exist $r$ and $(X,Y,Z)$ such that
$(x,y,z)=r(X,Y,Z)+b(Y,-X,0)$,
${x^2+y^2\over R^2}+{z^2\over r^2}=1$
and $X^2+Y^2+Z^2=1$, where $R=(r^2+b^2)^{1/2}$.}
\label{fig:loop}
\end{figure}

Define a weave $\Delta_t$ such that
the coordinates $\Delta_t^a$ of each point on the weave is
$\dot\Delta_t^a=t(t^{-1})^{a}_{\ b}\dot\Delta^b$,
where a dot means
the derivative with respect to a parameter of the loop and
$(t^{-1})^a_{\ b}$ and $t$ are the inverse and the determinant
of $t^a_{\ b}$ respectively.
We assume that $t^a_{\ b}$ (thus $(t^{-1})^a_{\ b}$) is uniform
over a single loop component of the weave since each
loop component is of a size of order $l_p$;
in other words, $t^a_{\ b}$ constructed from a classical
metric is considered to be insensitive
to the Planck scale structure of space-time.
Thus the weave is determined by
\begin{equation}
\Delta_t^a=t(t^{-1})^a_{\ b}\Delta^b,
\label{eq:delta}
\end{equation}
up to a loop-wisely undetermined additive constant.
We assume here that each of $\Delta_t^a$ and $\Delta^a$ is one of
the coordinates  of each point on a loop component relative to
the center of the loop component,
the positions of the centers of all the loop components are
unchanged so that the undetermined constants are zero,
and that $(t^{-1})^a_{\ b}$  is defined at the center of each loop
component.
The resulting weave, representing the Kerr-Newman metric,
is such that
\begin{equation}
\left[\matrix{\Delta_t^x\cr\Delta_t^y\cr\Delta_t^z\cr}\right]=
\left[\matrix{-XZ/\rho& Y/\rho& X\cr
              -YZ/\rho&-X/\rho& Y\cr
              \rho& 0& Z\cr}\right]
\left[\matrix{\alpha&0&0\cr 0&\alpha&0\cr 0&0&1\cr}\right]
\left[\matrix{\Delta^x\cr\Delta^y\cr\Delta^z\cr}\right].
\label{eq:weave}
\end{equation}
This transformation consists of two parts,
uniform expansions (or contractions)
of loop components in the direction perpendicular to
the $z$-direction and rotations of loop components
so that the non-expanded (or non-contracted) direction
coincides with $(X, Y, Z)$ at the center of
each loop component $(x,y,z)$,
(See Fig.\ref{fig:loop}).
The magnitude of the changes from the flat weave
depends on $\sqrt{x^2+y^2}$ and $z$.
This construction is well defined everywhere except where
$\Sigma=0$ or $r\le -M+\sqrt{M^2+e^2-b^2\cos^2\theta}$
as mentioned above.

Let us turn to discuss the meaning of the transformation.
We are interested in physical information contained
in the transformation and hence in the weave produced.
In particular, we examine the areas of the event hrizons
for different values of $M$, $e$ and $b$.
The area of the horizon computed from the metric
Eq~(\ref{eq:kerr}) or (\ref{eq:metric}) is $4\pi(r_+^2+b^2)$.
Remember that the area of a surface in the loop representation
is proportional to
the number of times loops intersect the surface.
By expanding or contracting the flat weave,
the number of intersections with the surface changes.
If the surface is closed, then an expansion or contraction
in the direction
transverse to the normal of the surface does not alter
the total number of intersections
but one in the normal direction does.
If it is expanded (contracted) by a factor,
the number of intersections
increases (decreases) by the same factor
since the weave consists of microscopic loops with a size
of order $l_p$ whose centers are distributed randomly
with an average number density of order $l_p^{-3}$.

First,
if $M=0$ and $e=0$, that is, $\alpha=1$, then $\Delta_t$ would be
another flat weave obtained from $\Delta$ by just changing
the orientations of the loop components regardless of the value
of $b$ but fixed.
The corresponding geometry is still flat up to the Planck scale
structure, which is undetermined from classical metrics.
This means that if we fix a surface, then the net change of the
number of times the loops intersect it is zero in macroscopical
sense, that is, the area of the surface does not change.
Therefore, the transformations with $\alpha=1$ keep
the classical flat geometry unchanged
although they change the Planck scale structure,
to which the classical geometry is insensitive.

Next, consider cases of static black holes, $b=0$ but $M\ne 0$
(the Schwarzschild metric for $e=0$ and the Reissner-Nordstrom
metric for $e\ne 0$).
In these cases, due to the presence of non-unit $\alpha>1$,
the number of times the loops intersect a surface does change.
However, the direction of expansion of each loop component is
rotated to be transverse to the radial direction at the center
of the loop component.
In other words, $(X, Y, Z)$ agrees with $(x/r, y/r, z/r)$.
Therefore, the total number of times the loop intersect
a spherical surface does not change.
Since $b=0$ (static black holes), the horizon is a spherical
surface at $r=r_+$; hence, the area of the horizon does
not change from the value in the flat geometry, which is
$4\pi r_+^2$.

The cases of rotating black holes, $b\ne 0$ and $M\ne 0$
(the Kerr metric for $e=0$ and the Kerr-Newman metric for
$e\ne 0$), are different.
The direction of the expansion of each loop component is
rotated not to be transverse to the radial direction at the center
of the loop component.
In other words, $(X, Y, Z)$ does not agree with $(x/r, y/r, z/r)$.
Furthermore, the horizon is not a spherical surface
but an elliptical one in this coordinate system.
Therefore, the number of times the loops intersect
the horizon may change; hence, its area may change from the value
at the same position in the flat geometry, which is
$4\pi(r_+^2+b^2)[{1\over2}+{1\over2}{r_+\over\sqrt{r_+^2+b^2}}
{\sinh^{-1}(b/r_+)\over b/r_+}]$.

More precise computation of the area of the horizon is as follows.
The area of an infinitesimal surface $dS_a(x)$ at $x$
by means of the black hole weave is
\begin{equation}
dA=\left\vert l_p^2\oint du\dot\Delta_t^a(u)\delta^3(x,\Delta_t(u))
dS_a(x)\right\vert
\end{equation}
up to a multiplicative constant.
This expression means that the area is the number of times
the loops in the weave intersect the surface $dS_a$.
Then, by inserting Eq~(\ref{eq:delta}),
rewrite it in terms of the flat weave
\begin{equation}
dA=\left\vert l_p^2\oint du\dot\Delta^b(u)\delta^3(x',\Delta(u))
t(t^{-1})^a_{\ b}dS_a(x)\right\vert,
\end{equation}
where $x'$ is the position transformed from $x$ by
$t(t^{-1})^a_{\ b}$.
{}From the definition of the flat weave $\Delta$,
this is simply the area of the surface $t(t^{-1})^a_{\ b}dS_a$
at $x'$ in the flat space,namely
\begin{equation}
dA=\left(\tilde{\tilde h}^{cd}(x')t(t^{-1})^a_{\ c}dS_a(x)
t(t^{-1})^b_{\ d}dS_b(x)\right)^{1/2}
\end{equation}
and hence it becomes
\begin{equation}
dA=\left(\tilde{\tilde q}^{ab}(x)dS_a(x)
dS_b(x)\right)^{1/2}
\end{equation}
by definition of the transformation $t^a_{\ b}$.
Therefore, by integrating over the surface of the event horizon
at $r=r_+$, we can find the area of it, namely
$A=4\pi(r_+^2+b^2)$.
Here, we have recovered the area of the event horizon
from the language of loops.

Note that in any case,
$\Delta_t$ coincides with the reoriented flat weave
with $b=0$ as $\sqrt{x^2+y^2+z^2}\to\infty$.
This fact implies that the black hole weaves are
``asymptotically flat."

\section{Remarks}\label{sec:remarks}

In this section we remark implications and applications of
the black hole weaves constructed in this paper.

(i)
The coordinate system we chosen is just a choice for convenience.
Another appropriate choice of coordinates
may give a different weave,
which is supposed to be related to ours by a diffeomorphism
up to  Planck scale structures.
Physical information invariant under diffeomorphisms
must be extracted
in an appropriate way in the loop representation
quantum gravity.

(ii)
The weaves themselves
cannot determine causal structures and
cannot distinguish black and white holes
(the change of the alternative signs
in $X$ and $Y$ is merely equivalent to
a rotation of coordinate system about $x$- or $y$-axis
by the amount of $\pi$ radian),
which are determined by canonical conjugate variables.
In the loop representation,
the canonical conjugate information
is coded in topological operations
of loops\cite{loop};
moreover, in the quantum theory,
the conjugate information is uncertain when the 3-metric is certain.
Therefore, we have to understand the physical meaning of
our construction of black holes
in the language of the loop representation quantum gravity.

(iii)
The construction of our weave is appropriate
only  where $q_{ab}$ is slowly varying with respect to $h_{ab}$,
that is, where $t\equiv\alpha$ is close to the unity.
Therefore, at small $r$ our weave may fail
to approximate the corresponding 3-metric.
In particular, at the singularlity and the region
$r\le -M+\sqrt{M^2+e^2-b^2\cos^2\theta}$
we do not consider its validity.

(iv)
Our construction of the black hole weaves is independent from
the choice of a flat weave.
That is, given a flat weave, one can determine the corresponding
black hole weave with the three papameters $M$, $e$ and $b$.
Therefore, the aspect of the metric the black hole weave
approximates is the same aspect of the flat metric
the flat weave approximates.
By choosing a flat weave with a wider range of aspects of the metric,
one can construct the corresponding black hole weave
with the wider range of aspects of the black hole metric.
This could be done by imposing more conditions on the flat weave
by means of more operators one is interested in
(i.e. the volume operator \cite{discrete}).

(v)
The weaves are constructed from classical metrics.
Therefore, they do not have information about
the Planck scale geometry at this stage.
However, they may have a potential to have further structure
at the Planck scale.
In this sense, we expect that the weave might provide
an aspect of ``precise" picture of the Planck scale geometry,
to which the classical picture of geometry is insensitive.

(vi)
A graviton theory on the flat weave, which describes
graviton physics in the flat metric space,
has been constructed by means of
the non-perturbative loop representation
quantum gravity \cite{mapM,gravitons}.
It may be interesting to construct a graviton theory on
the black hole weave constructed here,
which describes a graviton physics in the corresponding
black hole background metric,
and study the Hawking effect,
as suggested by Ashtekar \cite{abhay}.

(vii)
Another possible application might be to study the black hole
thermodynamics.
Since the weave might provide an aspect of ``precise" picture
of the Planck scale geometry, to which classical picture of
geometry is insensitive, if differential geometry of GR
describes the black hole ``thermodynamics," then
``loop geometry" of the loop representation
might describe a black hole ``statistical mechanics."

\vskip0.3cm
The author thanks Carlo Rovelli for helpful comments and discussions.

\end{document}